# Temperature Dependence Of Core Loss In Cobalt Substituted Ni-Zn-Cu Ferrites

A. Lucas [1,2], R. Lebourgeois [1], F. Mazaleyrat [2], E. Labouré [2,3]
1. THALES R&T, Campus Polytechnique, 1 av. Augustin Fresnel, 91767 Palaiseau, France
2. SATIE, ENS de Cachan, 61 av. du Président Wilson, 94235 Cachan, France
3. LGEP – SUPELEC, Plateau de Moulon, 11 rue Joliot-Curie, 91192 Gif Sur Yvette

**Abstract** : The temperature dependence of core loss in cobalt substituted Ni-Zn-Cu ferrites was investigated. $Co^{2+}$ ions are known to lead to a compensation of the magneto-crystalline anisotropy in Ni-Zn ferrites, at a temperature depending on the cobalt content and the Ni / Zn ratio. We observed similar behaviour in Ni-Zn-Cu and it was found that the core loss goes through a minimum around this magneto-crystalline anisotropy compensation. Moreover, the anisotropy induced by the cobalt allowed a strong decrease of core loss, a ferrite having a core loss of 350 mW/cm$^3$ at 80°C was then developed (measured at 1.5 MHz and 25 mT). This result represents an improvement of a factor 4 compared to the state of art Ni-Zn ferrites.

# 1 Introduction

Nickel-zinc-copper ferrites are essential materials because of their high permeability in MHz range. Moreover, their low sintering temperature makes them suitable for the realization of integrated components in power electronics. As for nickel-zinc ferrites, cobalt substitution is an efficient technique to decrease the permeability [1] and the magnetic losses of nickel-zinc-copper ferrites [2]. The effect of cobalt is to allow the pinning of the domain wall by inducing a magnetic anisotropy [3]. The consequence is a decrease of $\tan\delta_\mu$ at high frequency [4] and also an improvement of core loss [2]. The aim of this paper is to study the effect of cobalt substitution on core losses versus temperature. A lot of papers investigated core losses of spinel ferrites versus temperature but it was mostly for Mn-Zn power ferrites [5] and not at high frequency. There are very few papers concerning core losses versus temperature of Ni-Zn or Ni-Zn-Cu ferrites [6][7], moreover, they only investigated losses of a single composition measured at low frequency (50 kHz). In order to understand how the ferrite composition influences the core loss, ferrites with three Ni / Zn ratios were studied (Ni / Zn = 0.43, 1 and 3) with, for each ratio, cobalt substitutions up to 0.035 mol per formula.

# 2 Experimental procedure

## 2.1 Sample preparation

Ferrites were synthesized using the conventional ceramic route. The raw materials ($Fe_2O_3$, NiO, ZnO, CuO) were ball milled for 24 hours in water. $Co_3O_4$ was then added before the calcination around 800°C in air for 2 hours. The calcined ferrite powder was then milled by attrition for 30 min. The resulting powder was compacted using axial pressing. The sintering was performed at 935°C for 2 hours in air. Magnetic characterizations were done on ring shaped samples with the following dimensions : outer diameter = 6.8 mm; inner diameter = 3.15 mm; height = 4 mm.

## 2.2 Sample measurements

Bulk density was deduced from weight and dimensions. Saturation magnetization was measured on a magnetic balance. Initial complex permeability was measured versus frequency between 1 MHz and 1 GHz using an HP 4291impedance-meter. Static initial permeability ($\mu_s$) was defined as µ' at 1 MHz because for these ferrites µ' is constant from very low frequencies to the megahertz range. For the permeability versus temperature measurements, the rings were wound with a copper wire and placed in an oven going from –70°C to 150°C. $\mu_s$ was deduced from the inductance measured at 100 kHz by an 4194A impedance-meter Agilent. Core losses were measured at 1.5 MHz and 25 mT using a Clark-Hess 258 wattmeter and a 100 W Kalmus HF amplifier.

# 3 Results and discussion

## 3.1 Physicochemical characterizations

Three different series of ferrites were studied:

- $Ni_{0.24}Zn_{0.56}Cu_{0.20}Co_{\varepsilon}Fe_{1.98}O_{4-\gamma}$
- $Ni_{0.40}Zn_{0.40}Cu_{0.20}Co_{\varepsilon}Fe_{1.98}O_{4-\gamma'}$
- $Ni_{0.60}Zn_{0.20}Cu_{0.20}Co_{\varepsilon}Fe_{1.98}O_{4-\gamma''}$

Four formulations were done for each series, with cobalt rate of 0, 0.014, 0.028 and 0.035 mol per formula. The copper rate was 0.20 mol to allow the densification below 950°C. Three Ni / Zn ratios were studied in order to investigate the influence of cobalt substitution on ferrite with different magneto-crystalline anisotropy. Indeed, $K_1$ of the host crystal increases with nickel content [8].

The samples were sintered at 935°C for 2 hours in air. Figure 1 shows the X-ray diffraction pattern of a $Ni_{0.40}Zn_{0.40}Cu_{0.20}Fe_2O_4$ ferrite sintered at 935°C. Only the spinel structure can be observed, none of the precursor oxides are present in the sintered materials.

The bulk densities of the sintered ferrites are shown in figure 2. All the ferrites have a low porosity ($\rho$ > 92 % of the theoretical density). The cobalt substitution does not seem to have a significant effect on densification, a slight increase of the densification can however be noticed. This phenomenon is probably a consequence of the slight increase in the iron default of the ferrite.

Microstructure of the ferrites were also observed, SEM pictures are shown in figure 3. The grain size is not affected by the Ni / Zn ratio. Even if the nickel is known to lead to a more difficult densification [9], it is not observed here. The cobalt rate does not change the microstructure. All the ferrites exhibit the same kind of microstructure, with a low porosity and an average grain size between 1.5 and 2 µm. The grain size repartition is relatively homogenous, although one can see a few larger grains with defaults inside.

### 3.2 Magnetic characterisations

Saturation magnetization ($M_s$) at room temperature was measured (see table 1). This parameter mainly depends on the Ni / Zn ratio. $M_s$ increases up to a Ni / Zn ratio close to 1, then, it decreases with the zinc content. The behaviour of the developed ferrites is coherent with previous studies concerning Ni-Zn [8] and Ni-Zn-Cu [9] ferrites. The maximum of the saturation magnetization is obtained for the $Ni_{0.40}Zn_{0.40}Cu_{0.20}Co_{\varepsilon}Fe_{1.98}O_{4-\gamma'}$ ferrites. Copper slightly lowers $M_s$ compared to the corresponding Ni-Zn ferrite [10].

Cobalt substitution does not have a significant influence on the saturation magnetization. Cobalt introduction must lead to an increase of the saturation magnetization, due to $Co^{2+}$ ions magnetic moment, which is higher than the other divalent ions present in octahedral sites ($Ni^{2+}$ and $Cu^{2+}$) [11]. The measurement device used for this study is probably not precise enough to discern a variation in $M_s$ with such a small amount of $Co^{2+}$.

For each ferrite, the static initial permeability and the core loss were measured versus temperature. The results are shown in figure 4.

The initial static permeability versus temperature of the cobalt substituted ferrites is not monotonous ; A local maximum can be observed even if for high $\mu_s$ samples, it is flattened. This is due to the contribution of the cobalt ions to the magneto-crystalline anisotropy ($K_1$). Several studies were carried out on this phenomenon, which can be useful to increase the permeability around a particular temperature [12]. The permeability is due to two contributions : the domain wall displacements and the magnetization rotation :

$$\mu_s - 1 = \frac{3}{16} \cdot \frac{M_s^2}{K_1} \cdot D \quad \text{for the domain wall motion contribution [13]}$$

$$\mu_s - 1 \propto \frac{2}{3} \cdot \frac{M_s^2}{K_1} \quad \text{for the spin rotation contribution [14]}$$

With D the average grain size and $K_1$ the magneto-crystalline anisotropy of the ferrite. In this study, we consider that the grain size is similar whatever the composition, so this parameter does not affect the permeability. The variation of permeability versus temperature is then only depending on the variations of $M_s$ and $K_1$ versus temperature.

As the saturation magnetization of these ferrites is monotonically decreasing versus temperature, such a change in permeability is therefore due to a change in magnetic anisotropy. $K_1$ of the Ni-Zn-Cu spinels is weak and negative (between $-2$ and $-6,5.10^3$ J/m$^3$, for $NiFe_2O_4$, $ZnFe_2O_4$ and $CuFe_2O_4$ [8]). The cobalt ferrite has a high and positive $K_1$ (close to $300.10^3$ J/m$^3$ at room temperature [8]). Addition of a small amount of cobalt will thus lead to a compensation of $K_1$. Looking at the previous relations, if there is a magneto-crystalline anisotropy compensation, the permeability will go through a maximum. This explains the apparition of a local maximum in $\mu_s(T)$ curves of cobalt substituted ferrites at a compensation temperature ($T_0$) increasing with the cobalt content and $K_1$ of the Ni-Zn-Cu host crystal.

The compensation temperature of the three series of ferrites are shown in table 2, these results are established from $\mu_s(T)$ curves (Figures 4.A1, B1 and C1). For a given Ni /Zn ratio, $T_0$ increases with the cobalt content. For the same cobalt rate, $T_0$ increases with the nickel content, i.e. with magneto-crystalline anisotropy of the Ni-Zn-Cu host crystal. At low temperature, the magneto-crystalline anisotropy is positive due to $Co^{2+}$ ions contribution. At high temperature, Ni-Zn-Cu-host crystal contribution becomes preponderant and $K_1$ is negative. The measurements are thus coherent with the single-ion model theory.

The cobalt substitutions are also known to pin the domain walls and to lower the domain wall displacements contribution to the permeability. This results in a decrease of $\mu_s$ with the increase of the cobalt content, as can be seen in figures 4.A1 and 4.B1. For the third series (Ni / Zn = 3, figure 4.C1), the magneto-crystalline compensation is particularly marked. The permeability of cobalt substituted ferrites can then become higher than that of the cobalt-free ferrite (curve B, figure 4.C1). It seems that for these ferrites, a slight cobalt substitution leads to a smaller $K_1$ than that of the cobalt-free ferrite.

**Temperature dependence of core loss**

Core loss variations mainly depend on saturation magnetization and effective anisotropy (since we consider that the microstructures are similar for all the materials). Indeed, it is known that larger grain size can increase core loss [15]. For the cobalt-free ferrites these two parameters decrease versus temperature.

The $Ni_{0.24}Zn_{0.56}Cu_{0.20}Fe_{1.98}O_4$ ferrite (curve A of the figure 4.A2) has core loss decreasing until 20°C and rapidly increasing above 120°C. To understand this behaviour, we have to look at the temperature dependence of the permeability. Until 80°C, the anisotropy decreases faster than the saturation magnetization, which results in an increase of the permeability and a decrease of the core loss. Above 80°C, the initial static permeability reaches 1000, shifting the μ" resonance frequency towards 3 MHz. As the measurements are done at 1.5 MHz, relaxation losses become therefore predominant and lead to a fast increase of the core loss at high temperature.

When the nickel content increases, the Curie temperature rises. For the $Ni_{0.40}Zn_{0.40}Cu_{0.20}Fe_{1.98}O_{4-\gamma}$ ferrite (curve A, figure 4.B2), $T_c$ is around 340°C. The core loss measurements between −50 and 150°C are continuously decreasing. This is again related to the variation of permeability which constantly increases while maintaining a resonance frequency far enough from 1.5 MHz.

For the $Ni_{0.60}Zn_{0.20}Cu_{0.20}Fe_{1.98}O_4$ ferrites (curve A, figure 4.C2), one can note that the permeability is almost constant in the studied temperature range. The core losses are thus nearly stable between −50 and 150°C, a slight decrease can be observed at high temperature. Concerning the core loss values of the Ni-Zn-Cu ferrites, the higher the magneto-crystalline anisotropy, the higher the core loss. The power losses at room temperature increase from 600 mW/cm$^3$ to 2800 mW/cm$^3$ when nickel content goes from 0.24 mol to 0.60 mol per formula.

Cobalt substituted ferrites have a completely different behaviour. The core loss goes through a minimum around $T_0$ (magneto-crystalline compensation temperature) and their values are also highly reduced near this compensation compared to the cobalt-free ferrites. This phenomenon appears to be similar to the $K_1$ compensation observed in Mn-Zn spinels [12]. This compensation, due to $Fe^{2+}$ ions, also leads to a minimum in the core loss. If this is well known for Mn-Zn ferrites, it is the first time that it is observed in Ni-Zn based ferrites.

Such low core losses around $T_0$ are not only due to the $K_1$ compensation but are also the result of the cobalt induced anisotropy ($K_u$). Small amounts of $Co^{2+}$ ions in Ni-Zn spinels are known to pin the domain walls, however, the temperature dependence of this induced anisotropy is not well known. In order to study this parameter, Core loss versus magnetic induction of the $Ni_{0.40}Zn_{0.40}Cu_{0.20}Co_{0.028}Fe_{1.98}O_{4+\gamma}$ ferrite was measured at different temperatures (figure 5). The cobalt substituted ferrites have not a classical behaviour of core loss versus induction. When induction increases, the core loss increases almost linearly until a threshold induction corresponding to the required energy to unpin the domain walls. Above this threshold induction (which depends on $K_1$ and $K_u$), the anisotropy induced by the cobalt is no longer efficient so the core loss increases fast. This phenomenon appears clearly on figure 5 for the measurement at 20°C, the core loss increases slightly up to 35 mT and then rises rapidly. When the temperature increases, the threshold induction decreases and, for the measurement done at 100°C, the core loss nearly recovers a classical behaviour with an evolution proportional to the square of the magnetic induction. This ferrite has a $T_0$ close to 10°C, therefore the measurements done on figure 5 tend to prove that the cobalt induced anisotropy is more efficient around the magneto-crystalline anisotropy compensation.

Consequently, around $T_0$ the magnetic configuration is doubly favourable in order to have low power loss with a weak effective anisotropy and an induced anisotropy that improve the linearity of the ferrite. This can then lead to very low core loss of respectively 150 and 200 mW/cm$^3$ (at 1.5 MHz, 25 mT and 20°C) for Ni / Zn = 0.43 and Ni / Zn = 1 ferrites with cobalt content of 0.028 mol. The third series of ferrites have higher core loss with a minimum of 400 mW/cm$^3$. It seems that $K_1$ of the Ni-Zn-Cu host crystal (which increases due to high nickel content) is too high to have very low core loss.

## 3 Conclusion

The study of the temperature dependence of core loss of cobalt-substituted Ni-Zn-Cu ferrites highlighted that the core losses were minimum around the magneto-crystalline anisotropy compensation. This is extremely interesting for adapting the ferrite to the operating range temperature for power applications. Today, power electronics need materials that can work at high frequency and high temperature (> 80°C). The ferrites studied in this paper can perfectly answer to these issues. Indeed, for the $Ni_{0.40}Zn_{0.40}Cu_{0.20}Co_\varepsilon Fe_{1.98}O_{4+\gamma}$ ferrites (figure 4.B2), core loss at 80°C can be divided by a factor 4 thanks to cobalt substitution. These materials represent also a real improvement compared to the state of art Ni-Zn ferrite for radiofrequency, with core loss 4 times lower than a commercial ferrite with $\mu_s=120$ (figure 6).

**References**

[1]  T. Y. Byun, S. C. Byeon, K. S. Hong, "Factors affecting initial permeability of Co-substituted Ni-Zn-Cu ferrites", IEEE, vol 35, Issue 5, Part 2, pages : 3445-3447 (1999)

[2]  R. Lebourgeois, J. Ageron, H. Vincent and J-P. Ganne, low losses NiZnCu ferrites (ICF8), Kyoto and Tokyo, Japan (2000)

[3]  L. Néel, J. Phys. Radium 13, 249 (1952)

[4]  J. G. M. De Lau, A. Broese Van Groenou, J. De Phys. IV 38, C1-17 (1977)

[5] V. T. Zaspalis, E. Antoniadisa, E. Papazogloua, V. Tsakaloudia, L. Nalbandiana and C. A. Sikalidisb, "The effect of $Nb_2O_5$ dopant on the structural and magnetic properties of MnZn ferrites", Journal of Magnetism and Magnetic Materials Volume 250, Pages 98-109, (2002)

[6]  H. Su, H. Zhang, X. Tang, Y. Jing, Y. Liu, "Effects of composition and sintering temperature on properties of NiZn and NiCuZn ferrites" Journal of Magnetism and Magnetic Materials 310, p17–21, (2007)

[7]  Y. Matsuo, M. Inagaki, T. Tomozawa, and F. Nakao, IEEE Transactions On Magnetics, Vol. 37, n°. 4, (2001)

[8]  J. S. Smit and H. P. J. Wijn, Ferrites, Philips technical library (1961)

[9]  J. Ageron ,Thesis (1999)

[10]  G. K. Joshi, A. Y. Khot and S. R. Sawant, "Magnetisation, curie temperature and Y-K angle studies of Cu-substituted and non substituted Ni-Zn mixed ferrites", Solid State Communications, vol 65 n°12, 1593-1595 (1988)

[11]  E. Rezlescu, L. Sachelarie, P. D. Popa, and N. Rezlescu," Effect of Substitution of Divalent Ions on the Electrical and Magnetic Properties of Ni–Zn–Me Ferrites", IEEE transactions on magnetics, vol 36 n°6 (2000)

[12]  H. Pascard, Basic concepts for high permeability in soft ferrites, J. Phys. IV France, 377-384 (1998)

[13]  R. Lebourgeois, C. Le Fur, M. Labeyrie, M. Paté, J-P. Ganne, "Permeability mechanisms in high frequency polycrystalline ferrites", Journal of Magnetism and Magnetic Materials, Volume 160, Pages 329-332,(1996)

[14] J. L. Snoek, Physica 14, 207 (1948)

[15] R. Lebourgeois, S. Duguey, J.-P. Ganne, J.-M. Heintz, "Influence of $V_2O_5$ on the magnetic properties of nickel–zinc–copper ferrites", Journal of Magnetism and Magnetic Materials, Volume 312, Issue 2, Pages 328-330, (2007)

| Ferrites | | Ms (emu/g) |
|---|---|---|
| Ni/Zn=0.43 | Co=0 | 65.5 |
| Ni/Zn=0.43 | Co=0.014 | 65.5 |
| Ni/Zn=0.43 | Co=0.028 | 64.4 |
| Ni/Zn=0.43 | Co=0.035 | 65.2 |
| Ni/Zn=1 | Co=0 | 79.8 |
| Ni/Zn=3 | Co=0 | 72.4 |

Table 1 : Saturation magnetization at room temperature of materials sintered at 935°C for 2 hours in air.

| % Cobalt | $T_0$ Ni/Zn = 0.43 | $T_0$ Ni/Zn = 1 | $T_0$ Ni/Zn = 3 |
|---|---|---|---|
| 0.014 | -45°C | -50°C | -28°C |
| 0.028 | -7°C | 5°C | 30°C |
| 0.035 | 3°C | 17°C | 50°C |

Table 2 : Magneto-crystalline anisotropy compensation of the three series of ferrites, determined from $\mu_s(T)$ curves.

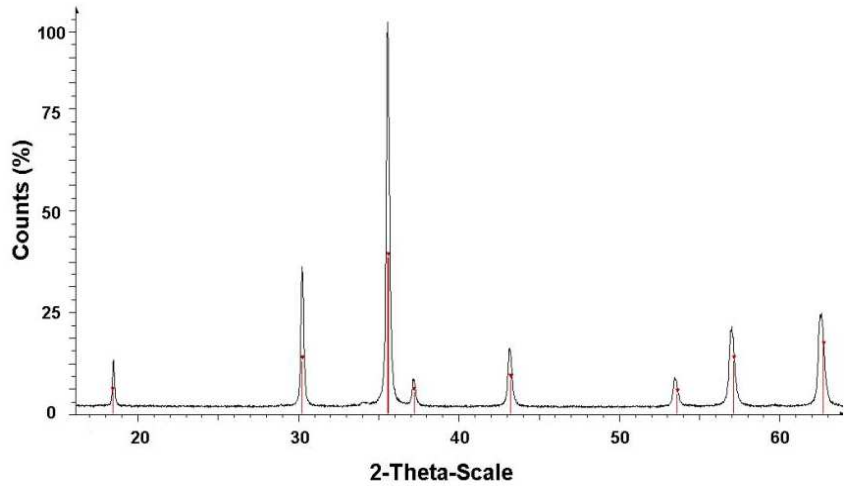

Figure 1 : X-ray diffraction pattern of $Ni_{0.40}Zn_{0.40}Cu_{0.2}Fe_{1.98}O_4$ ferrite (black curve), comparison with JCPDF 00-052-0277 (red peaks).

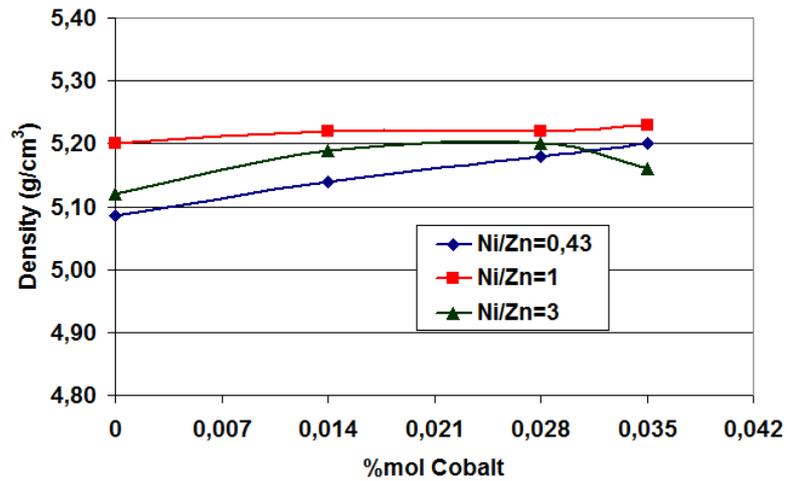

Figure 2 : Density of the three series of ferrites sintered at 935°C for 2 hours in air.

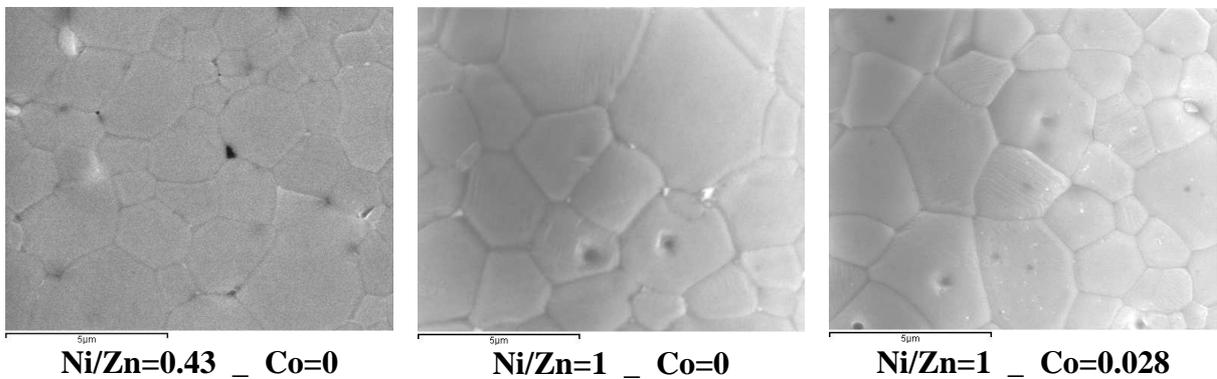

**Ni/Zn=0.43 _ Co=0**     **Ni/Zn=1 _ Co=0**     **Ni/Zn=1 _ Co=0.028**

Figure 3 : SEM Micrographics of three materials sintered at 935°C for 2 hours in air.

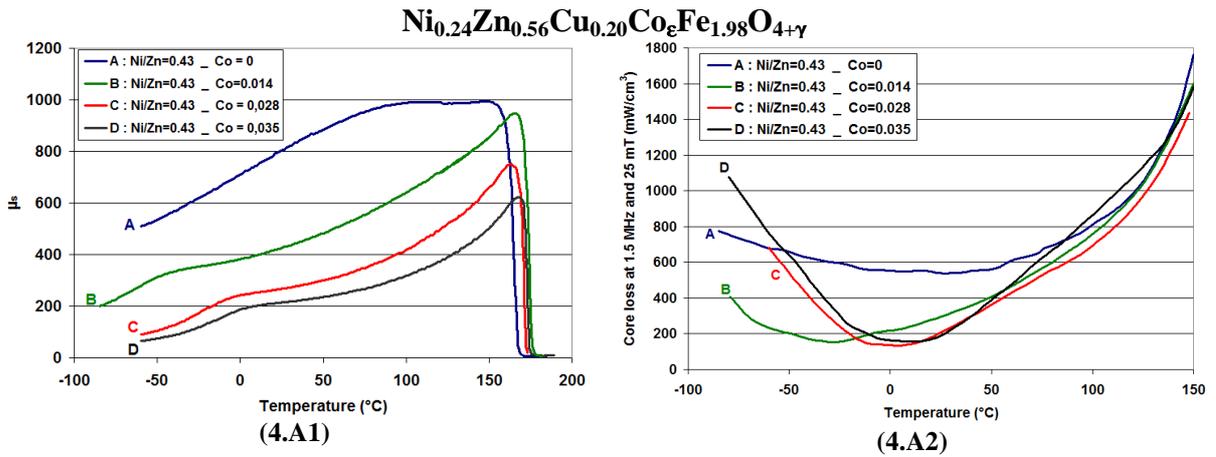

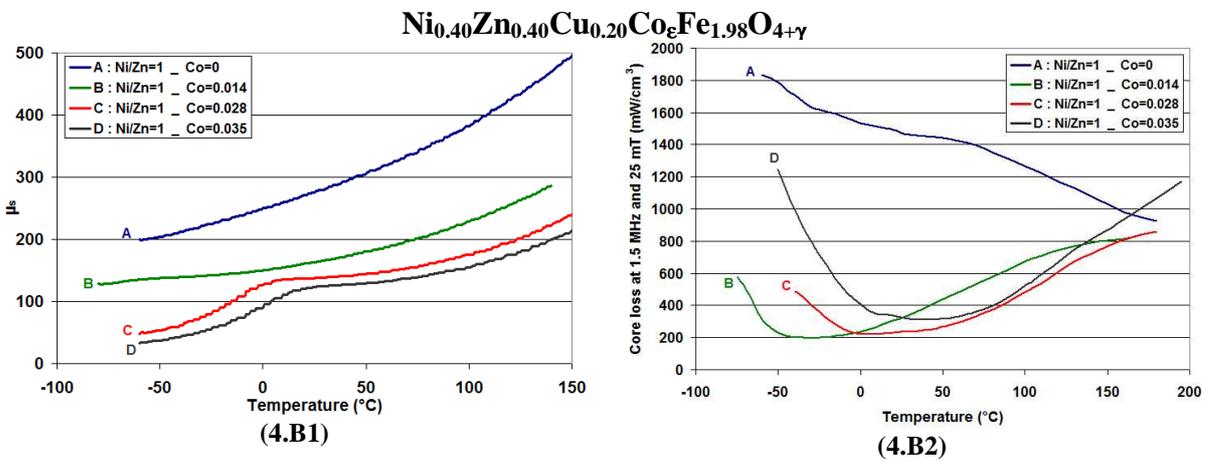

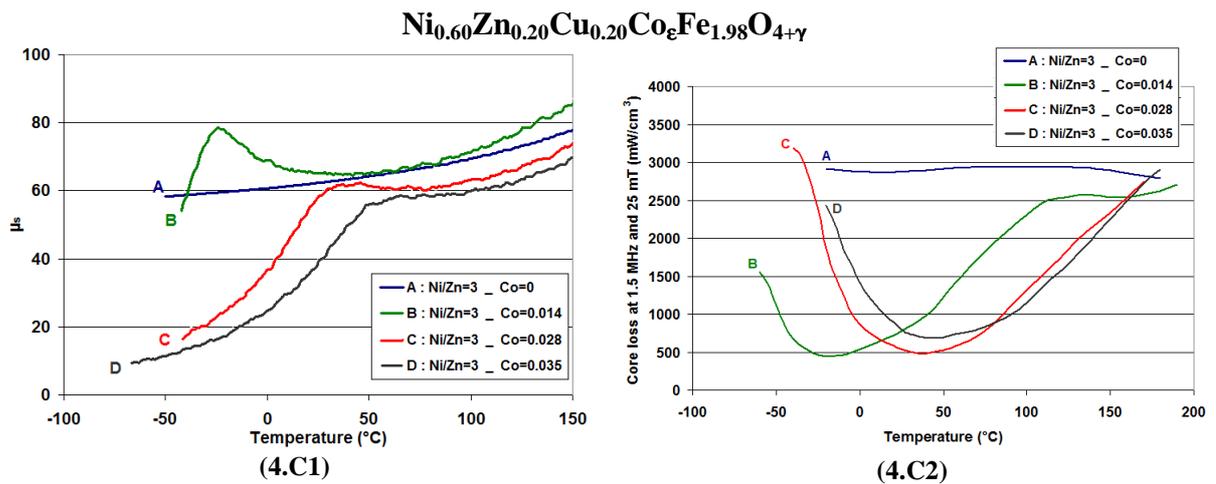

Figure 4 : Initial static permeability versus temperature and core loss at 1.5 MHz and 25 mT versus temperature of $Ni_{0.24}Zn_{0.56}Cu_{0.20}Co_{\varepsilon}Fe_{1.98}O_{4+\gamma}$, $Ni_{0.40}Zn_{0.40}Cu_{0.20}Co_{\varepsilon}Fe_{1.98}O_{4+\gamma}$ and $Ni_{0.60}Zn_{0.20}Cu_{0.20}Co_{\varepsilon}Fe_{1.98}O_{4+\gamma}$ ferrites

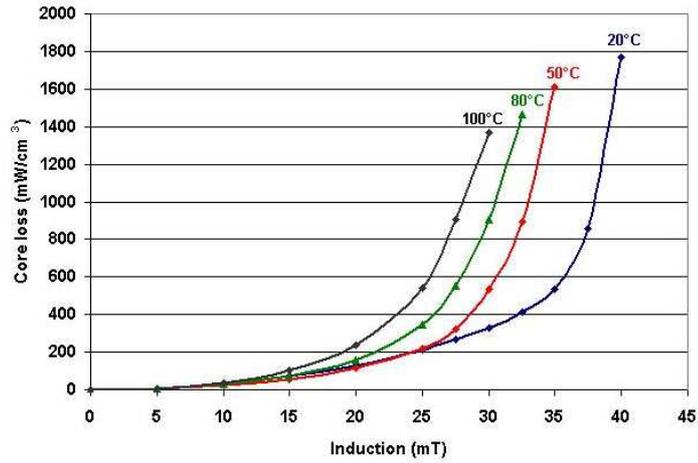

Figure 5 : Core loss of the $Ni_{0.40}Zn_{0.40}Cu_{0.20}Co_{0.028}Fe_{1.98}O_{4+\gamma}$ ferrite measured at 1.5 MHz versus magnetic induction for different temperatures.

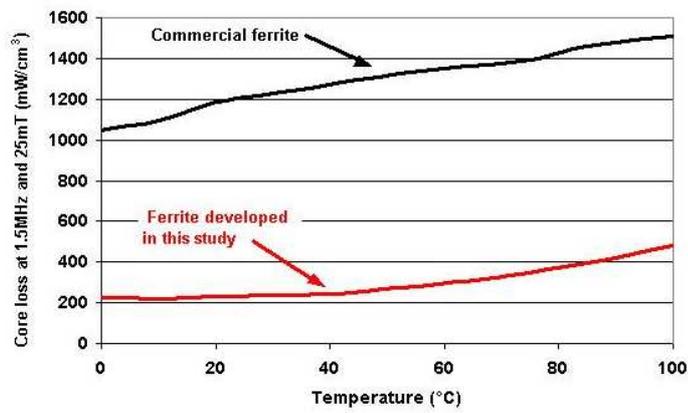

Figure 6 : Core loss versus temperature measured at 1.5 MHz and 25mT. Comparison between a commercial Ni-Zn ferrite ($\mu_s=120$) and the $Ni_{0.40}Zn_{0.40}Cu_{0.20}Co_{0.028}Fe_{1.98}O_{4+\gamma}$ ferrite ($\mu_s=130$)